# Crystal Structure Prediction via Particle Swarm Optimization

Yanchao Wang, Jian Lv, Li Zhu and Yanming Ma*

*State Key Laboratory of Superhard Materials, Jilin University, Changchun 130012, China*

We have developed a powerful method for crystal structure prediction from "scratch" through particle swarm optimization (PSO) algorithm within the evolutionary scheme. PSO technique is dramatically different with the genetic algorithm and has apparently avoided the use of evolution operators (e.g., crossover and mutation). The approach is based on a highly efficient global minimization of free energy surfaces merging total-energy calculations via PSO technique and requires only chemical compositions for a given compound to predict stable or metastable structures at given external conditions (e.g., pressure). A particularly devised geometrical structure factor method which allows the elimination of similar structures during structure evolution was implemented to enhance the structure search efficiency. The application of designed variable unit cell size technique has greatly reduced the computational cost. Moreover, the symmetry constraint imposed in the structure generation enables the realization of diverse structures, leads to significantly reduced search space and optimization variables, and thus fastens the global structural convergence. The PSO algorithm has been successfully applied to the prediction of many known systems (e.g., elemental, binary and ternary compounds) with various chemical bonding environments (e.g., metallic, ionic, and covalent bonding). The remarkable success rate demonstrates the reliability of this methodology and illustrates the great promise of PSO as a major technique on crystal structure determination.

# I. INTRODUCTION

Crystal structure occupies a central and often critical role in materials science, particularly when establishing a correspondence between material performance and its basic composition since properties of a solid are intimately tied to its crystal structure. Experimentally, structural determination through X-ray diffraction technique has been developed extremely well, leading to numerous crystal structures solved. However, it happens frequently that experiments fail to determine structures due to the obtained low quality X-ray diffraction data, particularly at extreme conditions (e.g. high pressure). Here, the theoretical prediction of crystal structures with the only known information of chemical composition independent on previous experimental knowledge is of greatly necessary. However, this is extremely difficult as it basically involves in classifying a huge number of energy minima on the lattice energy surface. Twenty years ago John Maddox even published an article in Nature to question the predictive power provided with only the knowledge of chemical composition.[1] His words still remain largely true, as evidenced by poor results of the latest blind test for crystal structure prediction[2].

Owing to significant progress in both computational power and basic materials theory, it is now possible to predict the crystal structure at zero Kelvin using the quantum mechanical methods, some of which are simulated annealing[3, 4], genetic algorithm[5-8], basin hopping[9, 10], metadynamics [11, 12] and data mining methods[13]. Simulated annealing, basin hopping, and metadynamics have focused on overcoming the energy barriers and can be successful when the starting structure is close to the global minimum, though it is not always a priori known. The data mining method relies heavily on the existence of an extensive database of good trial structures and is incapable of generating new crystal structure types in the absence of information on similar compounds. The genetic algorithm (GA) starts to use a self-improving method and is successful in accurately predicting many high pressure structures.[14-18]

We here have proposed a new methodology for crystal structure prediction based on the particle swarm optimization (PSO) technique within the evolutionary scheme.

With the PSO implementation, the computational expense of first-principles density functional calculations has been significantly reduced. This is stemmed from that PSO is a highly efficient global optimization method. We have successfully applied this method to the prediction of various known systems, such as elemental, binary and ternary compounds. The remarkable success rate demonstrates the reliability of this methodology and illustrates the great promise of PSO as a major tool on crystal structure determination.

This paper is arranged as follows. In Sec. II, the method and implementation of PSO algorithm will be discussed in details. A short overview of results obtained from our method is presented in Sec. III followed by the summary in Sec. IV.

## II. METHOD AND IMPLEMENTATION

PSO is a branch of evolutionary methodology, but quite different with GA. In particular, the major evolution operations of "crossover" and "mutation" in GA have been avoided. PSO was first proposed by Kennedy and Eberhart in the mid 1990s[19, 20]. As a stochastic global optimization method, PSO is inspired by the choreography of a bird flock and can be seen as a distributed behavior algorithm that performs multidimensional search. According to PSO, the behavior of each individual is affected by either the best local or the best global individual to help it fly through a hyperspace. Moreover, an individual can learn from its part experiences to adjust its flying speed and direction. Therefore, all the individuals in the swarm can quickly converge to the global position and near-optimal geographical position by the behavior of the flock and their flying histories. PSO has been verified to perform well on many optimization problems [21-25]. We have implemented PSO algorithm on crystal structure prediction in CALYPSO (Crystal structure AnaLYsis by Particle Swarm Optimization) code[26].

Our global minimization method through CALYPSO code for predicting crystal structures comprises mainly four steps as depicted in the flow chart of Fig.1: (1) generation of random structures with the constraint of symmetry; (2) local structural optimization; (3) post-processing for the identification of unique local minima by geometrical structure factor; (4) generation of new structures by PSO for iteration.

**Step 1: Generation of random structures with the constraint of symmetry**

Two types of variables are necessary to define a crystal structure: lattice parameters and atomic coordinates. There are six lattice parameters: three angles and three lattice vectors. Each atom has three coordinates coded as a fraction of the corresponding lattice vectors. The first step of our approach is to generate random structures symmetrically constrained within 230 space groups. Once a particular space group is selected, the lattice parameters are then confined within the chosen symmetry. The corresponding atomic coordinates are generated by the crystallographic symmetry operations through Matrix-Column pairs (W, w) [27] where the point operation W is a 3×3 matrix and the translation operation w is one column. Using the Matrix-Column pairs, one obtains new coordinates by matrix multiplication:

$$\begin{aligned}\tilde{x}_1 &= W_{11}x_1 + W_{12}x_2 + W_{13}x_3 + w_1\\ \tilde{x}_2 &= W_{21}x_1 + W_{22}x_2 + W_{23}x_3 + w_2\\ \tilde{x}_3 &= W_{31}x_1 + W_{32}x_2 + W_{33}x_3 + w_3\end{aligned}\ .$$

This can be written in an abbreviated form: $\tilde{x} = Wx + w$ . Within this Matrix-Column pairs operation, one random atomic coordinate can be used to generate other symmetrically related coordinates.

The generation of random structures ensures unbiased sampling of the energy landscape. The explicit application of symmetric constraints leads to significantly reduced search space and optimization variables, and thus fastens global structural convergence. For example, in the case of monoclinic crystals, the symmetric constraint limits the range of fractional atomic coordinates within 0 - 0.5, namely only half of the search space, at the same time, the optimization variables are reduced to five thanks to a fixed lattice angle ($90^0$). Moreover, we have applied a symmetry checking technique, in which the appearance of identical symmetric structures is strictly forbidden. This allows the generation of diverse structures, which are crucial for the efficiency of global minimization.

Since crystal structure prediction is performed on a blind base. The choice of the simulation cell sizes (not prior known) is critical to target the global minimum structure. In practice, one can choose all possible cell sizes to perform the separate simulations and then compare all the resulting structures to derive the global stable structure. However, this procedure is extremely computational costly and for some particular cases, is not affordable. Here, the fast learning ability of PSO technique has allowed us to implement a variable cell size technique enabling the intelligent selection of the correct cell sizes during the structural evolution, and thus the computational cost has been significantly reduced.

**Step 2: Local optimization**

The potential energy surface can be regarded as a multi-dimensional system of hills and valleys with saddle points connecting them. The valleys are the local basin of attractions on the potential energy surface. The local optimization (such as line minimization, steepest descents, conjugate gradient algorithm or Broyden-Fletcher-Goldfarb-Shanno algorithm) can drive the structural energy to the local minimum, which may or may not be the global minimum. The approach of locally optimizing every candidate has been used with great success. Local optimization increases the cost of each individual, but very effectively reduces the noise of the landscape, enhances comparability between different structures, and provides locally optimal structures for further use. We use free energy as fitness function throughout the simulation. Both the atomic coordinates and lattice parameters are locally optimized. Among the locally optimized structures, a certain number of worst ones are rejected, and the remaining structures participate in creating new structures through PSO for the next generation.

**Step 3: Post-processing for the identification of unique local minima**

Our method which solves the packing problem contains a critical step where a large number of preliminary trial structures are generated and then structurally optimized. At this step, many newly generated structures are very similar or even identical. The direct use of these similar structures to generate next generation will significantly slow down the convergence of structural search. It is thus highly beneficial to remove these duplicates to accelerate the search process. We have designed a method to identifying structural similarity named as geometrical structure factor on the basis of inter-atomic distances, which are calculated according to the so called “bond-types”. For example, if there are two types (A and B) of atoms in the simulation cell, three bond-types, i.e., A-A, A-B and B-B, will be evaluated. Then, one can determine the distances and number of the first and second nearest-neighbors for different bond-types, which are then listed in the matrices as backup for future comparison. Once a new structure is generated, the geometrical structure factor is applied to check the similarity of this structure with those in the saved matrices within specified tolerances. Specifically, if this structure shares the same number of bonds with one structure in the matrices, the deviation of bond length is then calculated according to the equation $\Delta \mathrm{d} = \sqrt{\sum_{i,j}(L_i - L_j^{'})^2 \delta_{i,j}}$, where $L_i$ and $L_j^{'}$ are the bond lengths in the two structures, respectively, and $\delta_{i,j}$ is the delta function. If the deviation ($\Delta d$) is less than the preset threshold, the two structures are considered to be equivalent. Thus, the newly generated structure will be discarded. Otherwise, it is kept and documented in the matrices. The matrices containing all the structure information are updated after local optimization at every generation and used in next generation.

**Step 4: Generation of new structures by PSO**

In the next generation, a certain number of new structures are generated by PSO. Within the PSO scheme, a structure (an individual) in the searching phase space is regarded as a particle. A set of individual particles is called a population or a generation. During the evolution equation (1) is used to update the positions of particles.

$$x_{i,j}^{t+1} = x_{i,j}^{t} + v_{i,j}^{t+1} \tag{1}$$

It is necessary to note that the velocity plays an important role on determination of the speed and direction of particle movement. The new velocity of each individual *i* at the *j*th dimension is calculated based on its previous location ($x_{i,j}^{t}$) before optimization, previous velocity ($v_{i,j}^{t}$), current location ($pbest_{i,j}^{t}$) with an achieved best fitness of this individual, and the population global location ($gbest^{t}$) with the best fitness value for the entire population according to equation (2).

$$v_{i,j}^{t+1} = \omega v_{i,j}^{t} + c_1 r_1 (pbest_{i,j}^{t} - x_{i,j}^{t}) + c_2 r_2 (gbest_{i,j}^{t} - x_{i,j}^{t}) \quad (2)$$

where $j \in \{1,2,3\}$, $\omega$ denotes the inertia weight, $c_1$ is self-confidence factor and $c_2$ is swarm confidence factor, $r_1$ and $r_2$ are two separately generated random numbers and uniformly distributed in the range [0, 1]. The velocity update formula includes random parameters ($r_1$ and $r_2$) to ensure good coverage of the searching space and avoid entrapment in local optima. As shown in equation (2), it is quite obvious that the movement of particles in the search space is dynamically influenced by their individual past experience ($pbest_{i,j}^{t}, v_{i,j}^{t}$) and successful experiences attained by the whole swarm ($gbest^{t}$). Thus the velocity makes the particles to move towards to global minimum and accelerates the convergence speed. Moreover, to overcome explosion and divergence, the magnitudes of the velocities are necessary to be confined within the range of [-0.1, 0.1].

In order to improve the efficiency of the procedure, only the low energy structures which are on the most promising area of the configuration space, are selected to produce the next generation. If the energy deviation between one particular structure and the current global structure is lower than the preset energy tolerance, this structure is then used to generate the next population by PSO, otherwise it is discarded. In order to keep the population diversity, a certain number of structures whose symmetries must be distinguished from any of previously generated ones, are generated randomly.

## III. APPLICATION AND RESULTS

Here we benchmark our methodology on systems with known structures. All the calculations were performed in the framework of density functional theory within the all-electron projector augmented wave method as implemented in the VASP code[28]. Some basic parameters used in CALYPSO code can be found in Table I. An overview of the benchmark systems including elements, binary compounds and ternary compounds with known structures can be found in Tables II and III.

### 3.1 Elements

Lithium is a "simple" metal at ambient pressure, but exhibits complex phase transitions under compression. Experimentally, it has been demonstrated that lithium takes the phase transition sequence of bcc→fcc→ hR1→cI16[29-32], above which new phases are observed but remain unsolved. In theory, the complex post-cI16 structures above 70 GPa, such as *Cmca*-24, *C*2, *Aba*2, *and P*42/*mbc*, are proposed by using molecular dynamics, genetic algorithm and "AIRSS" method, respectively [33-36]. Here, we used the PSO method for structure prediction through CALYPSO code to predict the stable structures at 0, 10, 40, 70, 80, 100, 120 and 300 GPa, and found all the experimental and theoretical structures mentioned above at certain pressure ranges. It is remarkable that for all the simulation of these complex structures, only less than 300 generated structures are needed to derive the correct results. For example, the cI16 structure is successfully identified at the 6$^{th}$ generation with a population size of 30, i.e., only 210 structures are generated and locally optimized. This illustrates the great efficiency of our methodology.

Other elements, such as carbon[37, 38], silicon[39-45] and magnesium[46, 47] (Table II), were also tested and the simulations quickly reproduced all the experimental structures. Particularly, several meta-stable structures [Fig. 2 ] proposed earlier by other theoretical methods [7, 48] of carbon were predicted at 0 GPa and the bc8 structure (mata-stable phase) of sillicon is also predicted at 2 GPa. This indicates that our method can be used to predict metastable structures.

### 3.2 binary compounds

Silica is a binary semiconductor which exhibits many novel polymorphs at elevated pressures. We have successfully reproduced the experimental α-quartz, stishovite, $CaCl_2$-type, α-$PbO_2$-type, and pyrite-type structures [49-53] at the certain pressure regimes by the structural prediction through CALYPSO code. Again, all the structures rapidly converge to the global minimum with less than 300 local optimizations. Specifically, the structural search easily found the α-quartz structure at the 5$^{th}$ generation with only 120 structures at 0 GPa. The history plot of the simulations by CALYPSO code for silica at 70 GPa is shown in Fig. 3(a). The experimental $CaCl_2$-type structure was found at the 5$^{th}$ generation.

Other binary compounds[26, 54-60] are also benchmarked as listed in Table III. All the simulations show fast convergence to the experimental structures.

### 3.3 ternary compounds

We have successfully identified the most stable structures of $MgSiO_3$ and $CaCO_3$ under pressure. The post-perovskite *Cmcm* structure of $MgSiO_3$[61] was quickly found in the 5$^{th}$ generation at 120 GPa with only 100 local optimizations. In addition, the metastable perovskite phase of $MgSiO_3$ was identified in the 6$^{th}$ generation with less than 120 structures. This simulation further illustrates that our method is able to find both stable and meta-stable structures. Moreover, the experimental calcite phase of $CaCO_3$ [62] has been reproduced in the 13$^{th}$ generation at 0 GPa. In Fig. 3(b), we show the history plot of the structural search on $CaCO_3$ with the only input information of chemical composition. At the 13$^{th}$ generation the enthalpy shows a pronounced drop, and the examination of lowest enthalpy structure confirmed the convergence to the experimental calcite structure [Fig. 3(b)].

## IV. CONCLUSION

We have developed a systemic methodology for the crystal structure prediction based on PSO technique within evolutionary scheme as implemented in CALYPSO code. Our method could efficiently search the free energy space of the lattice geometry and atomic configuration of a solid looking for the ground-state and meta-stable structures in complex systems. The key elements of the proposed approach are PSO algorithm, the state-of-art *ab initio* structural optimization based on density function theory, the symmetry constraint on the structural generation, and the geometrical structure factor technique on elimination of the similar structures. We have implemented a variable cell size technique enabling the intelligent selection of the correct cell sizes, and thus significantly reducing the computational cost. This methodology has been successfully applied to various known experimental structures on elemental, binary and ternary compounds with, but not limited to, metallic, ionic, and covalent bonding. Our method is proved to be powerful with high efficiency and high success rate. Future development of this highly efficient PSO technique on prediction of much larger systems (say ~100 atoms/cell or above) is foreseeing feasible and thus the predictive power on structure solutions of nanomaterials, surface or thin films, and bio-materials are highly expected. Within this PSO algorithm the doors towards materials design (e.g., design of novel superconductive, thermoelectric, superhard, and energetic materials, etc) could open.

## Acknowledgement

The authors acknowledge funding from the National Natural Science Foundation of China under grant No. 10874054 and the research fund for excellent young scientist in Jilin University (No. 200905003).

*To whom any correspondence should be addressed: mym@jlu.edu.cn

## Table and Figure captions

**TABLE I.** Some of standard input parameters of CALYPSO code.

**TABLE II.** Systems on elements with known structures on which calculations were performed by CALYPSO. All the experimental structures are reproduced within the given generations and population sizes. Note that our choice of population sizes is based on experience and it is highly possible that the use of smaller sizes could also result in the correct structures, with lower computational cost and less generated structures.

**TABLE III**. Systems on binary and ternary compounds with known experimental structures on which calculations were performed by CALYPSO. All the experimental structures are reproduced within the given generations and population sizes. Note that our choice of population sizes is based on experience and it is highly possible that the use of smaller sizes could also result in the correct structures, with lower computational cost and less generated structures.

**FIG. 1.** The flow chart of CALYPSO

**FIG. 2. (color online)** The meta-stable structures of carbon predicted by CALYPSO. (a) bc8 structure predicted at 2000 GPa. (b) $C_6$ *Im-3m* structure predicted at 0 GPa. (c) β-Sn structure predicted at 0 GPa. (d) Chiral framework $P6_122$ or $P6_522$ structure predicted at 0 GPa.

**FIG. 3.(color online)** Main figures: enthalpy history of CALYPSO structure search on $SiO_2$ (a) at 70 GPa and $CaCO_3$ (b) at 120 GPa. Inserts: the predicted $CaCl_2$ (*Pnnm*)

structure of $SiO_2$ (a) at the 5$^{th}$ generation and perovskite structure of $CaCO_3$ (30 atoms/cell) (b) at the 13$^{th}$ generation.

# Table I

| Parameters | Value |
|---|---|
| Minimal interatomic distances | 0.8 (Å) |
| Proportion of the structures generated by PSO | 0.6 |
| Precision of *k*-points sampling | 0.04-0.1 |
| Population size | 20-50 |
| Self confidence factor ($c_1$) | 2.0 |
| Swarm confidence factor ($c_2$) | 2.0 |
| Inertia weight ($\omega$) | 0.5 |

# Table II

| Systems | Pressure (GPa) | Structures | Generation | Population size |
|---|---|---|---|---|
| Li | 0 | Bcc[a] | 1 | 30 |
| | 0 | 9R[b] | 3 | 30 |
| | 10 | Fcc[c] | 1 | 20 |
| | 40 | hR1[d] | 4 | 30 |
| | 70 | cI16[d] | 7 | 30 |
| C | 0 | Graphite[e] | 30 | 30 |
| | 0 | Diamond[f] | 6 | 30 |
| Si | 2 | Bc8[g] | 6 | 30 |
| | 10 | cd[h] | 1 | 20 |
| | 10 | sh[i] | 2 | 20 |
| | 10 | β- Sn[j] | 3 | 20 |
| | 10 | *Imma*[k] | 4 | 20 |
| | 40 | *Cmca*[l] | 2 | 20 |
| | 40 | Hcp[l] | 4 | 20 |
| | 80 | Fcc[m] | 1 | 20 |
| Mg | 0 | Hcp[n] | 6 | 30 |
| | 100 | Bcc[o] | 4 | 30 |

[a]Reference 29. [b]Reference 30. [c]Reference 31. [d]Reference 32.
[e]Reference 37. [f]Reference 38. [g]Reference 39. [h]Reference 40.
[i]Reference 41. [j]Reference 42. [k]Reference 43. [l]Reference 44.
[m]Reference45. [n]Reference 46. [o]Reference 47.

# Table III

| System | Pressure (GPa) | Structures | Generation | Population size |
|---|---|---|---|---|
| $SiO_2$ | 0 | α-quartz[a] | 5 | 20 |
| | 20 | Stishovite[b] | 1 | 20 |
| | 70 | $CaCl_2$-type[c] | 5 | 30 |
| | 100 | α -$PbO_2$-type[d] | 4 | 20 |
| | 500 | Pyrite-type[e] | 15 | 20 |
| SiC | 0 | Zinc blende[f] | 6 | 30 |
| | 0 | Moissanite[f] | 3 | 30 |
| | 150 | Rock salt[g] | 2 | 30 |
| ZnO | 12 | Rock salt[h] | 2 | 30 |
| $TiH_2$ | 0 | *I4/mmm*[i] | 2 | 20 |
| | 0 | Fm-3m[j] | 3 | 20 |
| $MoB_2$ | 0 | R-3m[k] | 1 | 30 |
| $TiB_2$ | 0 | $AlB_2$-type[l] | 1 | 30 |
| $MgSiO_3$ | 120 | *Cmcm*[m] | 5 | 20 |
| $CaCO_3$ | 0 | Calcite[n] | 13 | 30 |

[a]Reference 49. [b]Reference 50. [c]Reference 51. [d]Reference 52.

[e]Reference 53. [f]Reference 54. [g]Reference 55. [h]Reference 56.

[i]Reference 57. [j]Reference 58. [k]Reference 59. [l]Reference 60.

[m]Reference 61. [n]Reference 62.

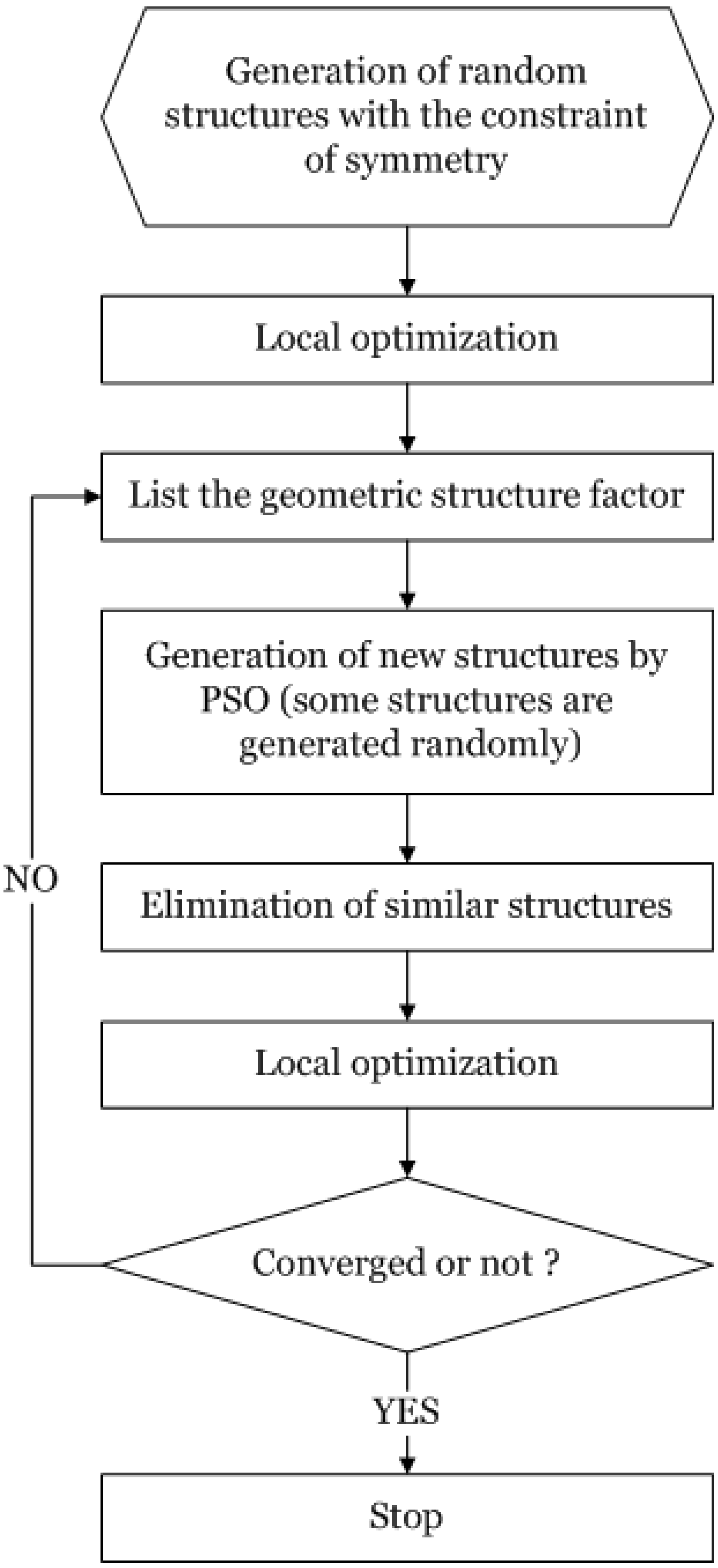
Generation of random structures with the constraint of symmetry
Local optimization
List the geometric structure factor
Generation of new structures by PSO (some structures are generated randomly)
Elimination of similar structures
Local optimization
Converged or not ?
NO
YES
Stop

**FIG. 1.**

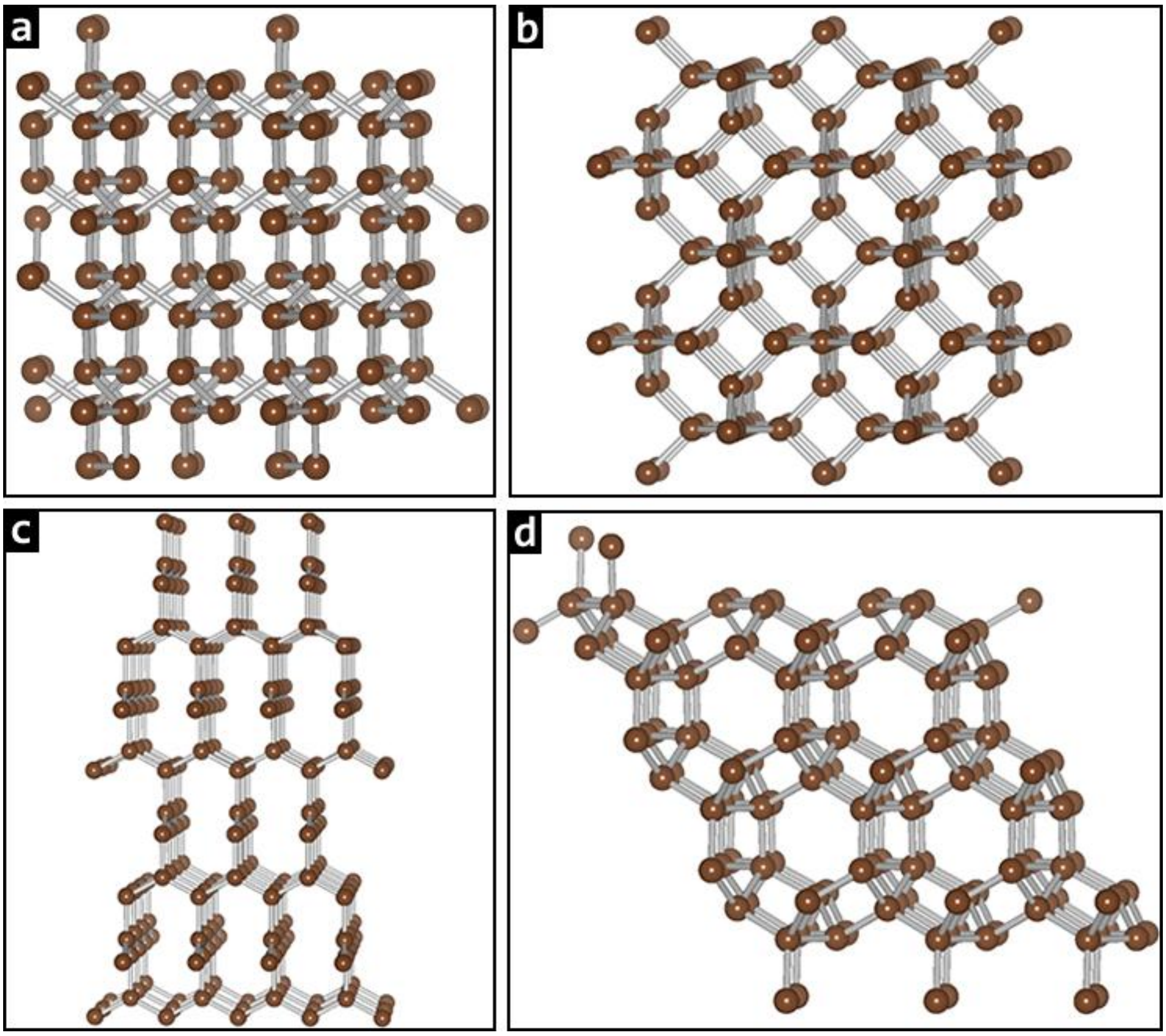

FIG. 2

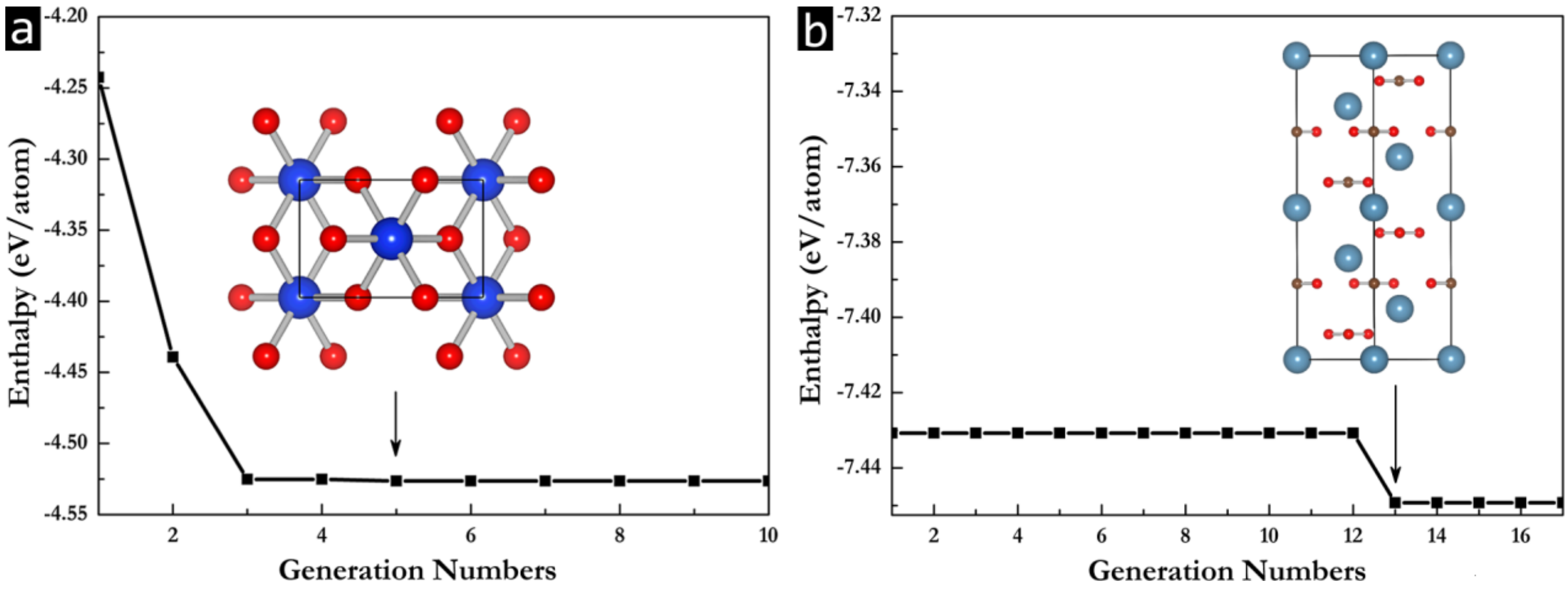

**FIG. 3**